\documentclass[pre,showpacs,showkeys,superscriptaddress,preprintnumbers,floatfixdvipdfmx]{revtex4}

\usepackage{amsmath} 
\usepackage{amssymb} 
\usepackage{amsfonts} 
\usepackage{graphicx}

\usepackage{url}
\usepackage{amsfonts,amsmath,amssymb,amsbsy,latexsym}

\begin{document}
\markboth{Noise-induced Statistical Periodicity and Almost Cyclic Sets in Random Lasota-Mackay maps}
{Sato, Padberg-Gehle, Iwata}


\title{Noise-induced Statistical Periodicity in Random Lasota-Mackey Maps}

\author{Yuzuru Sato}
\email[Electronic address: ]{ysato@math.sci.hokudai.ac.jp}
\affiliation{RIES/Department of Mathematics, Hokkaido University, 
N20 W10, Kita-ku, Sapporo 001-0020, Japan}
\affiliation{London Mathematical Laboratory, 8 Margravine Gardens,  London, W6 8RH, UK}

\author{Kathrin Padberg-Gehle}
\affiliation{
Institute of Scientific Computing, Technische Universit\"at Dresden, 
D-01062 Dresden, Germany}
\affiliation{
Institute of Mathematics and its Didactics, Leuphana Universit\"at L\"uneburg, Universit\"atsallee 1,
D-21335 L\"uneburg, Germany}


\begin{abstract}
Noise-induced statistical periodicity in a class of one-dimensional maps is studied. We show the existence of statistical periodicity in a modified Lasota-Mackey map and describe the phenomenon in terms of almost cyclic sets. A transition from a stable state to a periodic state of the density depending on the noise level is observed in numerical investigations based on trajectory averages and by means of a transfer operator approach. We conclude that the statistical periodicity is the  origin of the almost periodicity in noise-induced order.
\end{abstract}



\maketitle



Interaction between deterministic chaos and stochastic noise has been an important problem of statistical/nonlinear physics. Studies on  deterministic dynamical systems subjected to noise has been motivated by studies on stochastic dynamics, dynamical systems with a large degrees of freedom, non-autonomous dynamical systems, and applications in prediction and control of those systems. The central problem is to determine in which way the asymptotic behaviour of the system is affected by noise, and how much the original behaviour is altered. The simplest class of examples to treat this problem is that of one-dimensional maps stochastically perturbed by noise. In general, addition of noise to strongly chaotic hyperbolic dynamical systems will not result in any significant alternation of the statistical properties of the system \cite{Kifer74}. However, in other cases, even in one-dimensional maps, various noise-induced phenomena, such as noise-induced chaos \cite{Mayer-Kress81, Crutchfield82} or noise-induced order \cite{Matsumoto83} have been observed at the trajectory scale. The physical measures may be drastically altered by a small perturbation to those systems. They have been studied in terms of  the stability of invariant densities, and of the motion of invariant densities, which is on a measure for trajectories at a fixed time averaged over all possible noise realisations  \cite{Lasota87}. Such statistical behaviour can be characterised by investigating not only the fixed point of perturbed Perron-Frobenius operators, but also dynamical solutions of them. In this paper, we give a spectral decomposition analysis of highly non-trivial statistical periodicity with an example as a modified model given by Lasota and Mackey \cite{Lasota87}. The stochastic phenomena exhibited here, which are bifurcation-like phenomena in density space, have not been previously reported in low-dimensional random dynamical systems.

An example of affine random dynamical systems which shows a non-trivial noise-induced phenomenon in the density scale is given as the following random affine map \cite{Keener80, Lasota87, Lasota91}: 
\begin{equation}
 x_{n+1}=\alpha x_n+\lambda+\theta\xi_n ~~~(\mbox{mod} ~1), 
\label{eq:lm}
\end{equation}
where $\alpha<1$, and $\lambda, \theta \in [0,1]$. $\xi_n$ are random numbers uniformly distributed on $[0,1]$. 
This map is also known as the Nagumo-Sato map, a model of neurons \cite{nagumo1972response} when the stochastic term vanishes. In this deterministic case, it is known that there exists an uncountable set $\Lambda$ such that for any $\lambda\in\Lambda$, the orbit $\{x_n\}$ is aperiodic \cite{Keener80}. For the perturbed and unperturbed Lasota-Mackey map, the transfer operator describing the time evolution of densities allows us to apply the spectral decomposition theorem. As a result, the operator may have multiple spectra, and the densities may exhibit periodic limiting behaviour. In summary, adding noise to a deterministic map including aperiodic limiting solutions leads to periodic limiting behaviour of densities \cite{Lasota87}. This phenomenon is called {\it asymptotic periodicity.}  However, in these studies, with respect to the physical measure, and  with the given parameters $\alpha=1/2, \lambda=17/30$, the observed dynamics is a limit cycle with period 3, and the limiting behaviour of the density shows period 3, as well. Thus, this asymptotic periodicity describes simply a noisy limit cycle without qualitative  structural change of dynamics. 

We extend (\ref{eq:lm}) to a nonlinear random dynamical systems in the following nonlinear form: 
\begin{equation}
 x_{n+1}=\alpha x_n+\lambda-\frac{1}{1+e^{-\beta (\alpha x_n+\lambda-1)}}
+\theta\xi_n  \equiv S(x_n)+\theta\xi_n.
\label{eq:mlm}
\end{equation}
The parameter $\beta\in [0, \infty)$ introduces a non-affine structure to the Nagumo-Sato map, and this extension is known as a nonlinear model of neurons in the deterministic limit \cite{Aihara90}. The original model by Lasota-Mackey  (\ref{eq:lm}) is recovered when $\beta\rightarrow \infty$. We call this class of random dynamical systems {\it random Lasota-Mackey map}. 

\begin{figure}[htbp]
\begin{center}
\includegraphics[scale=0.5]{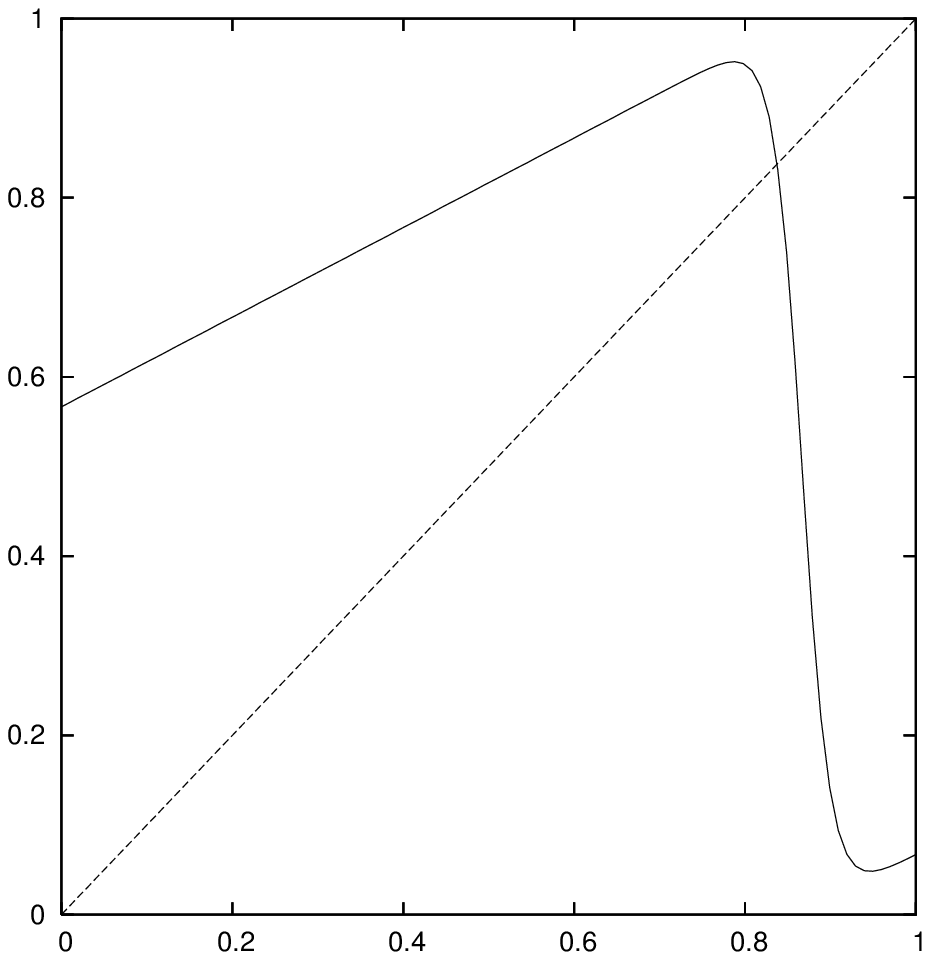}
\includegraphics[scale=0.55]{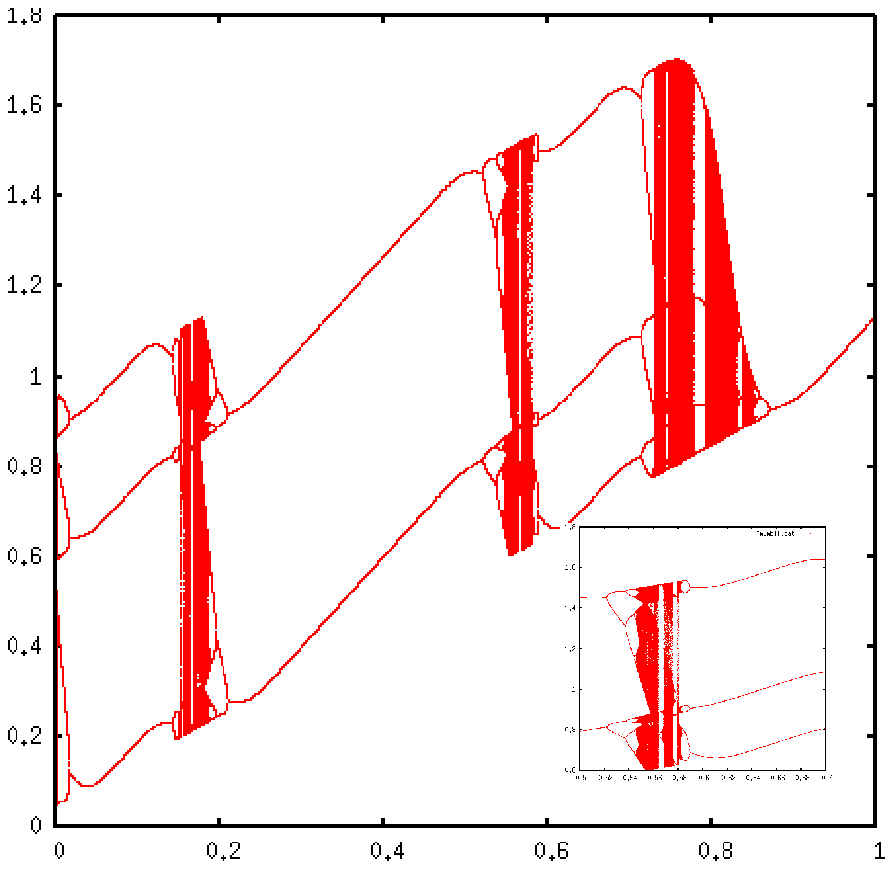}
\end{center}
\caption{Random Lasota-Mackey map with $\alpha=1/2, \lambda=17/30, \beta=120$ (left) and its bifurcation diagram varying the shift $b$ as in $S(x)+b$ (right). With $b=0$ and $\beta=120$ the dynamics is chaotic, with $\beta\rightarrow\infty$ the map has a stable period 3 limit cycle.}
\label{fig:map}
\end{figure}

To investigate the limiting behaviour of the random Lasota-Mackey map (\ref{eq:mlm}), we numerically observe the time-evolution of the densities based on an ensemble of underlying trajectories with many initial conditions and many noise realisations. 
In our numerical computation, densities are approximated as histograms with intervals with width $10^{-3}$ made from $10^6$ sample trajectories. We then numerically compute eigenfunctions and eigenvalues of the perturbed Perron-Frobenius operators  \cite{dellnitz99} and study these phenomena theoretically. For all numerical experiments, the parameters of the map are chosen to be $\alpha=1/2, \lambda=17/30, \beta=120$, and the amplitude of noise $\theta$ is varied independently. $\xi_n$ are random numbers uniformly distributed on $[0,1]$. The initial density $f_0$ is fixed to the uniform distribution on $[0,1]$. 
Our numerical experiments suggest that the modified Lasota-Mackey map, with parameters $\alpha=1/2, \lambda=\frac{17}{30}, \beta=120$, and $\theta\rightarrow 0$, is chaotic and has a stable invariant density. Adding noise, we observe clear statistical periodicity with $0.001<\theta<0.15$. Thus, we can say that there exists a deterministic chaotic map with an asymptotically stable invariant density, such that the addition of noise results in periodic behaviour of the densities on the sample measure in  average. We can also say that we present a more natural and non-trivial example of noise-induced statistical periodicity, compared to the one originally given by Lasota and Mackey \cite{Lasota87}. Varying the noise amplitude $\theta$, noise-induced statistical periodicity is observed in the random  Lasota-Mackey map, which we will describe below. 
\vspace{5mm}

\begin{figure}[htbp]
\begin{center}
\includegraphics[scale=0.3]{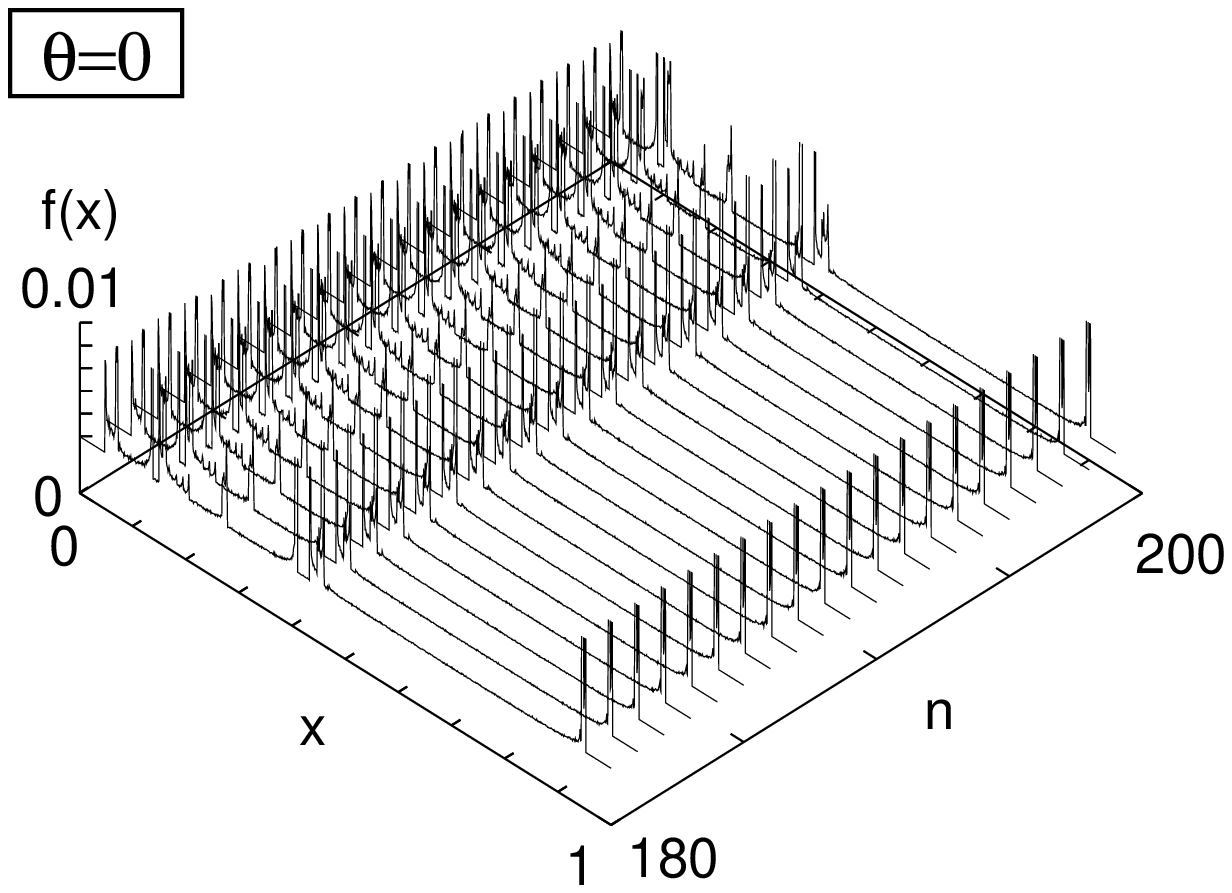}
\includegraphics[scale=0.3]{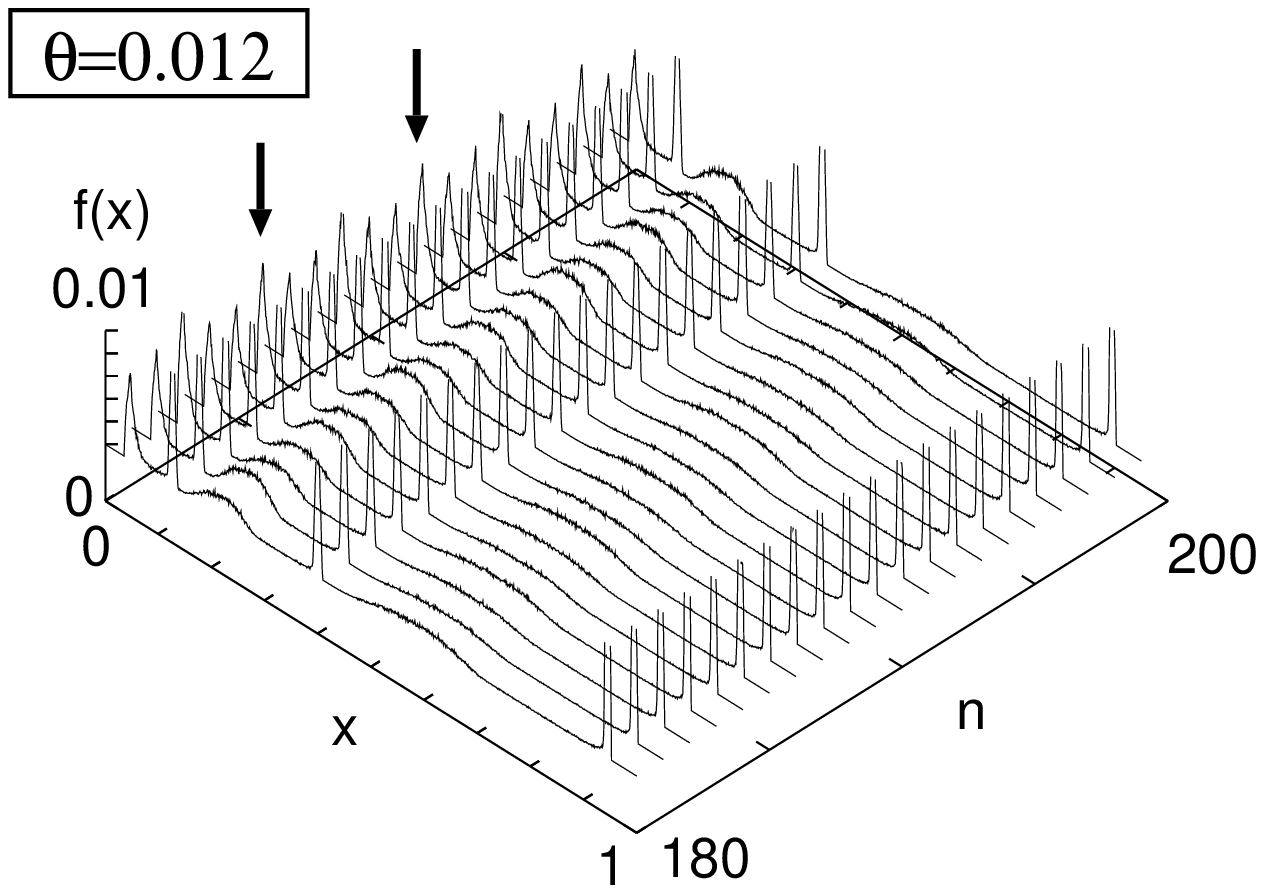}
\includegraphics[scale=0.3]{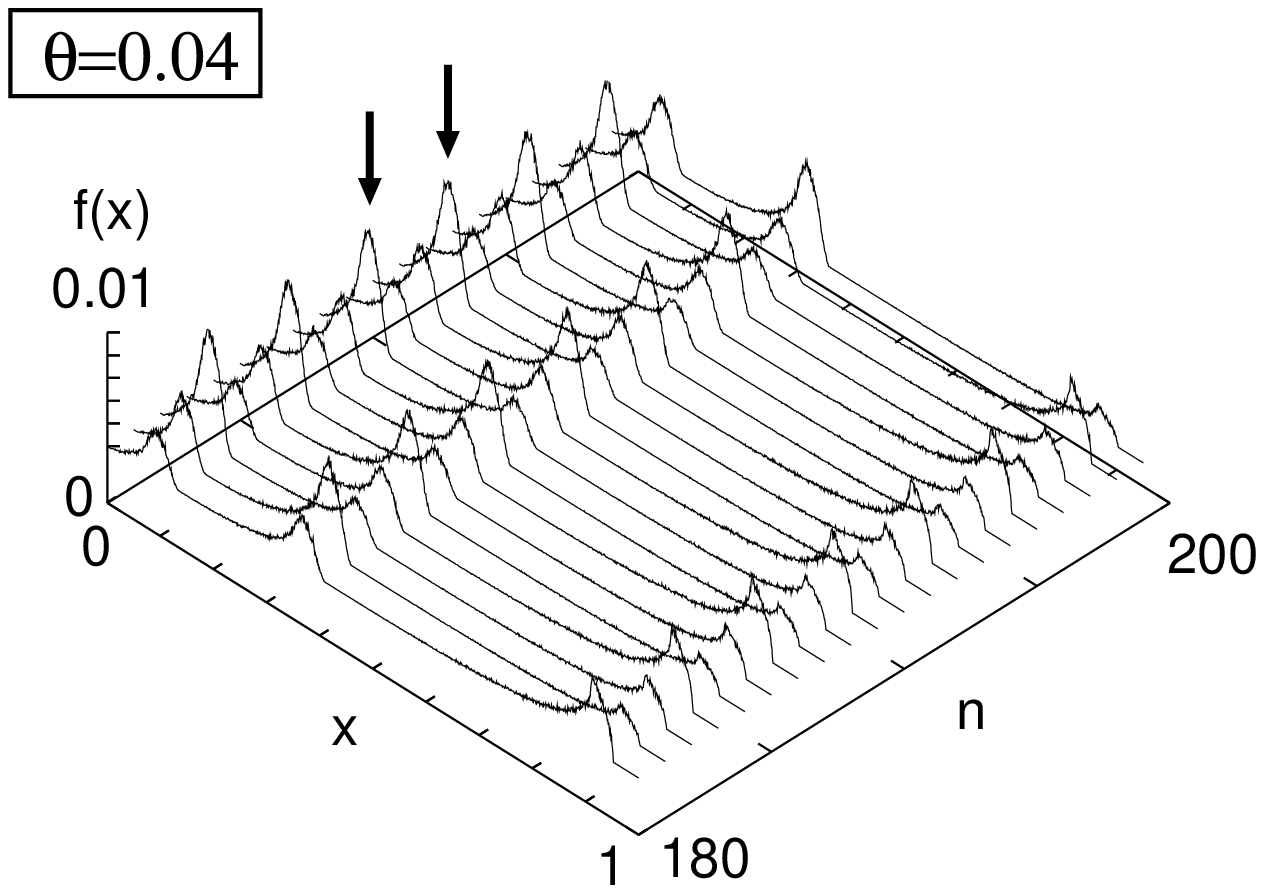}
\includegraphics[scale=0.3]{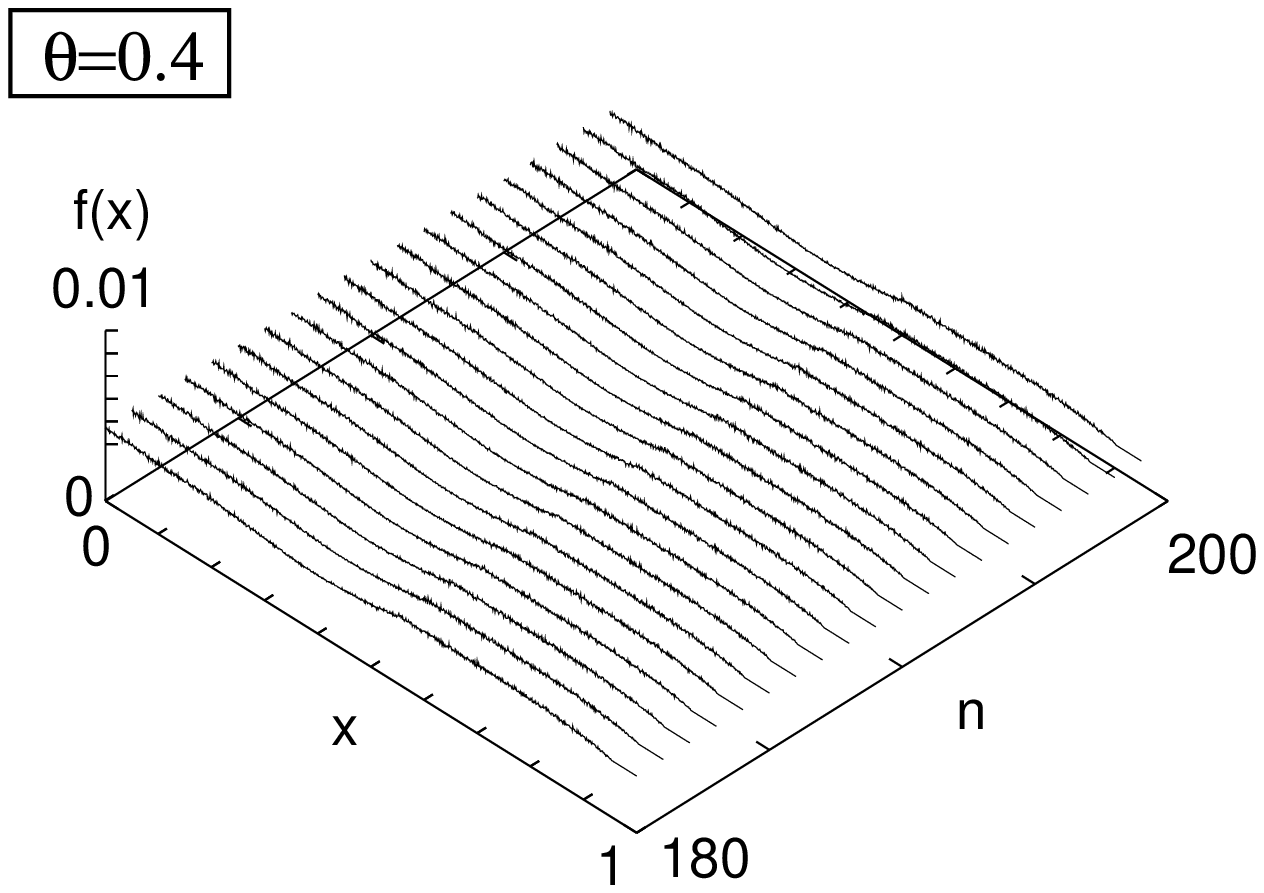}
\end{center}
\caption{Time evolution of invariant densities on the sample measure averaged over noise realisations. The parameters are $\alpha=1/2, \lambda=17/30, \beta=120$. The noise amplitude $\theta$ is varied as $\theta=0, 0.012, 0.04, and 0.4$.}
\label{fig:invdensity}
\end{figure}


Any linear operator $P: L^1({\bf R})\rightarrow L^1({\bf R})$ that satisfies 
\begin{equation}
P^n f(x)\ge 0, \mbox{\rm and} \int_{\bf R} P^n f(x)dx=\int_{\bf R} f(x) dx
\end{equation}
for a function $f(x)\ge 0$, is called Markov operator. The evolution of densities for a dynamical system $S$ can be described by a Markov operator $P$. We assume that $f(x)$ is a probability density whose integral can be normalised to $1$, and that the orbit $\{f_n\}=\{P^n f_0\}$ starts with a initial density $f_0$. For a deterministic dynamical system $x_{n+1}=T(x_n)$ defined on ${\bf R}$ additively perturbed by stochastic noise distributed over ${\bf R}$ with the density $h(x)$.  $P$ is given by a perturbed Perron-Frobenius operator \cite{Lasota87}, 
\begin{equation}
P f(x)=\int_{\bf R} h(x-T(y)) f(y) dy .
\end{equation}
The operator $P$ may have a stationary density $f^*$ which satisfies $P f^*(x)=f^*(x)$. 
Asymptotic periodicity is a type of limiting behaviour that can be observed in one-dimensional maps. When $P$ is a Markovian operator, the Komornik-Lasota spectral decomposition theorem  \cite{Komornik87} gives that for any $f_0$, we have 
\begin{equation}
P f_0(x)=\sum_{i=1}^r \mu_ig_i(x)+Qf_0(x), 
\end{equation}
where $g_1(x), g_2(x),\ldots, g_r(x)$ are the eigenfunctions of $P$, satisfying $g_i(x)g_j(x)=0$ if $i\neq j$ and $P g_i=g_{\sigma(i)}$, where $\sigma(i)$ is a permutation on $\{1, \ldots, r\}$. The absolute values of the corresponding eigenvalues $\gamma_1,\gamma_2,\ldots,\gamma_r$ are all $1$, and the support of $g_i$ is disjointed. The normalisation coefficients $\mu_i$ are determined depending on the initial density $f_0$ and satisfy $\sum_i\mu_i=1$. The operator $Q$ is called transient operator which satisfies $|| Q^n f_0||\rightarrow 0$ as $n\rightarrow \infty$. For the limiting behaviour, as the transient operator term vanishes, we have 
\begin{equation}
P^n f_0(x)\sim\sum_{i=1}^r \mu_{i}g_{\sigma^n(i)}(x), ~~(n\rightarrow \infty)
\label{eq:limitperiodic}
\end{equation}
When $r\ge 2$, the limiting behaviour of the density $f$ is called {\it asymptotically periodic} with period up to $r!$.
For the special case of $r=1$, the density approaches the fixed point $f^*$ with iterated application of $P$, and we have the invariant density,  
\begin{equation}
P^n f_0(x)\sim 1\cdot g_1(x)\equiv f^*(x), ~~(n\rightarrow \infty). 
\label{eq:limitstable}
\end{equation}
In this case, we call $f$ is {\it asymptotically stable}. 

As for our random Lasota-Mackey map $S$, the limiting behaviour of the density can be asymptotically periodic if $h(x)\ge 0$, and is asymptotically stable if $h(x) > 0$ a.e. The proof is given by the inequality $|S(x)|\le \alpha|x|+\lambda+1$, similarly to theorem 1 and 2 in \cite{Lasota87}. 
When $\theta$ is near $0$, the dynamics (\ref{eq:mlm}) is almost deterministic. The density is close to the invariant density of the deterministic chaotic dynamical system $x_{n+1}=S(x_n)$, and is asymptotically stable. Varying $\theta$ large, the dynamics becomes stochastic chaos \cite{chekroun2011stochastic, faranda2017stochastic} and the density can be asymptotically periodic. When $\theta$ is very large, the dynamics is almost stochastic as it is dominated by the stochastic dynamics $x_{n+1}=\theta\xi_n$. The density $f$ follows the external noise density $h(x)$, and is asymptotically stable again. It is easy to see that when $\theta$ is very large, the limiting behaviour of the density is asymptotically stable. 

If the asymptotic dynamics of our map $S$ is considerably restricted in the bounded region $[b,c] ~(b\le0,1\le c)$, and the support of $h(x)$ is included in $[b,c]$, then we have that 
\begin{equation} 
g_i\le P^n f(x)\le \frac{1}{\theta}\int_{[b,c]} P^{n-1}f(y)dy\le\frac{1}{\theta}
\end{equation}
for $x \in [b,c]$, where $\theta$ is the amplitude of the uniform distribution of the additive noise.  When $P^n f$ is asymptotically periodic, since each $g_i$ has disjoint support, we have 
\begin{equation}
r=\int_{[b,c]} \sum_{i=1}^r g_i(x)dx\le\int_{[b,c]}\frac{1}{\theta}dx=\frac{c-b}{\theta}.
\end{equation}
Thus, the upper bound of the period of the density is given by $O(\frac1\theta!)$, that means that the smaller noise induces the longer period. 

When the support of the invariant density is jointed, there is an unique eigenvalue $1$ and the system is asymptotically stable. However, when we have eigenvalues whose absolute value is very close to 1, other than the top eigenvalues, we have "almost disjointed supports" of density along with very narrow channels of probability flow which connect clusters of trajectories with each other. In such case, the system shows a slow oscillatory relaxation to the fixed point of the transfer operator. When this relaxation is significantly slow based on coexistence of multiple eigenvalues whose  absolute value close to 1, it can be called {\em statistical periodicity} in a phenomenological sense. Note that it does not satisfy the original mathematical conditions of asymptotic periodicity as a limiting behaviour. The invariant density is asymptotically stable in theory, but in finite time, we effectively observe  asymptotic periodicity. As our examples clearly show, it is expected to be robustly and dominantly observed in a broad class of random dynamical systems, and it is an evident noise-induced ordered phenomenon having emerged from a stable chaotic attractor. We claim that this mechanism underlies the origin of the order in noise-induced order \cite{Matsumoto83}, which is a noise-induced behaviour in a noised chaotic attractor with the negative Lyapunov exponent and the associated  almost periodicity.

To characterise the observed periodic behaviour in finite time , we use a mathematical concept termed {\it almost cyclic sets}  \cite{dellnitz99}. We say that a set $A$ is $\delta$-almost invariant with respect to a probability measure $\rho$ if
\begin{equation}
\frac{\rho(S^{-1}(A)\cap A)}{\rho(A)}=\delta
\end{equation}
Thus, $\delta$ can be viewed as the probability that points in $A$ are mapped into $A$ under $S$. In particular, if $A$ is an invariant set ($S^{-1}(A) = A$), then $\delta = 1$. 
If $S$ shows asymptotic periodicity, the notion of  $\delta$-almost cyclic sets with period $r$ describes this nearly periodic behaviour:
\begin{equation}
\frac{\rho(S^{-r}(A)\cap A)}{\rho(A)}=\delta
\end{equation}
In particular, for almost cyclic sets $A_0, \ldots, A_{r-1}$ it holds that \begin{equation}
  \frac{\rho(S^{-1}(A_{k+1 \mod r})\cap A_k)}{\rho(A_k)}\geq \delta,  
\end{equation} i.e.\ the different sets are approximately mapped into each other.
Almost-cyclic behaviour can be extracted from leading eigenfunctions $g_i$, $i=1,\ldots, r$ of the transfer operator corresponding to eigenvalues close to the $r$th roots of unity. In particular, the resulting $r$ almost cyclic sets are described by the supports of the $g_i$'s \cite{dellnitz99}.   

 When the system has a stable invariant set and $\delta$-almost cyclic sets with perturbation size $\theta$, we observed {\em noise-induced statistical periodicity} in finite time. The average probability flow rate through narrow channels among the clusters are characterised by $\delta$.


\begin{figure}[htbp]
\begin{center}
\includegraphics[scale=0.3]{denstraj0.012.eps}
\includegraphics[scale=0.3]{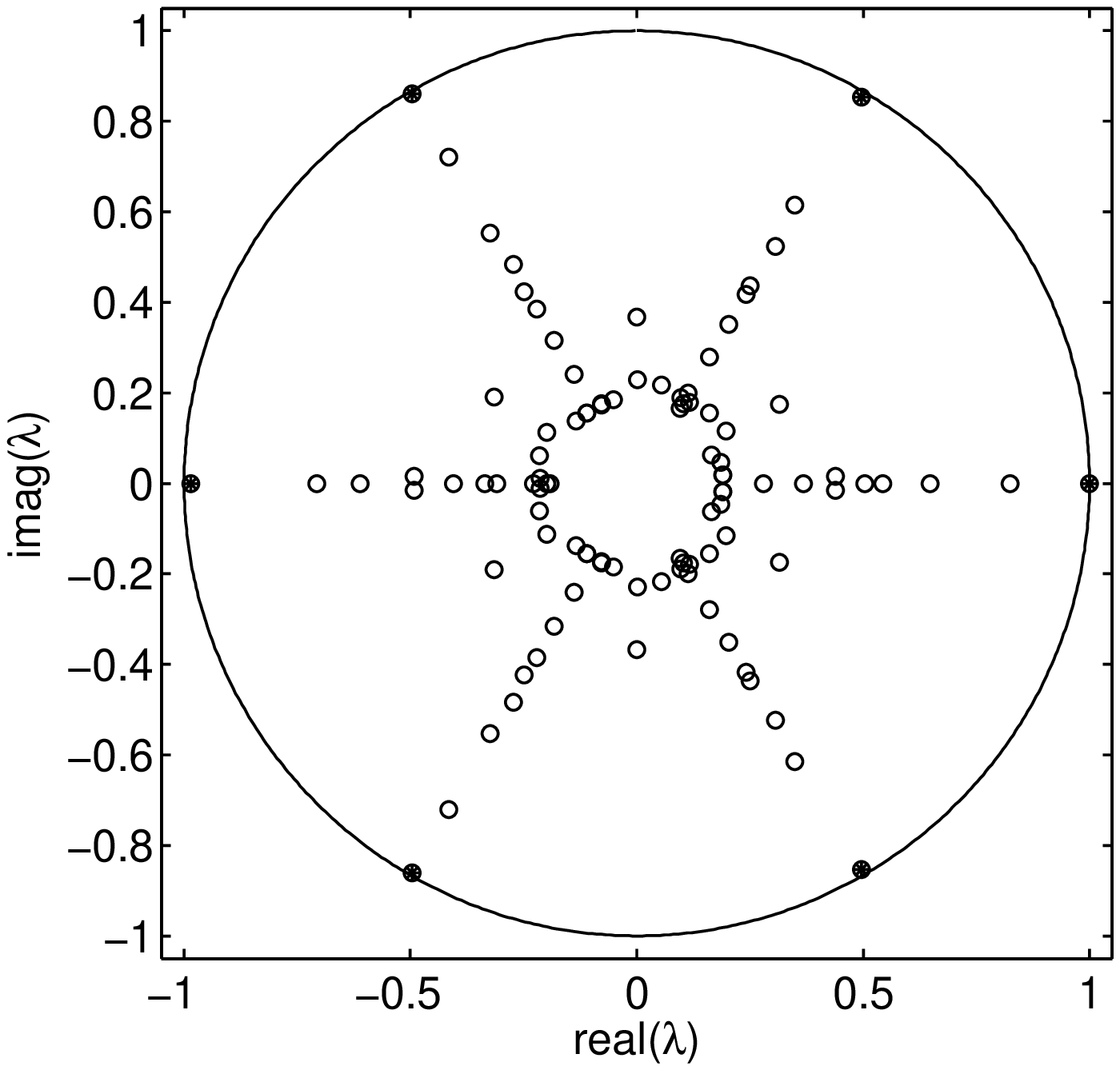}
\includegraphics[scale=0.3]{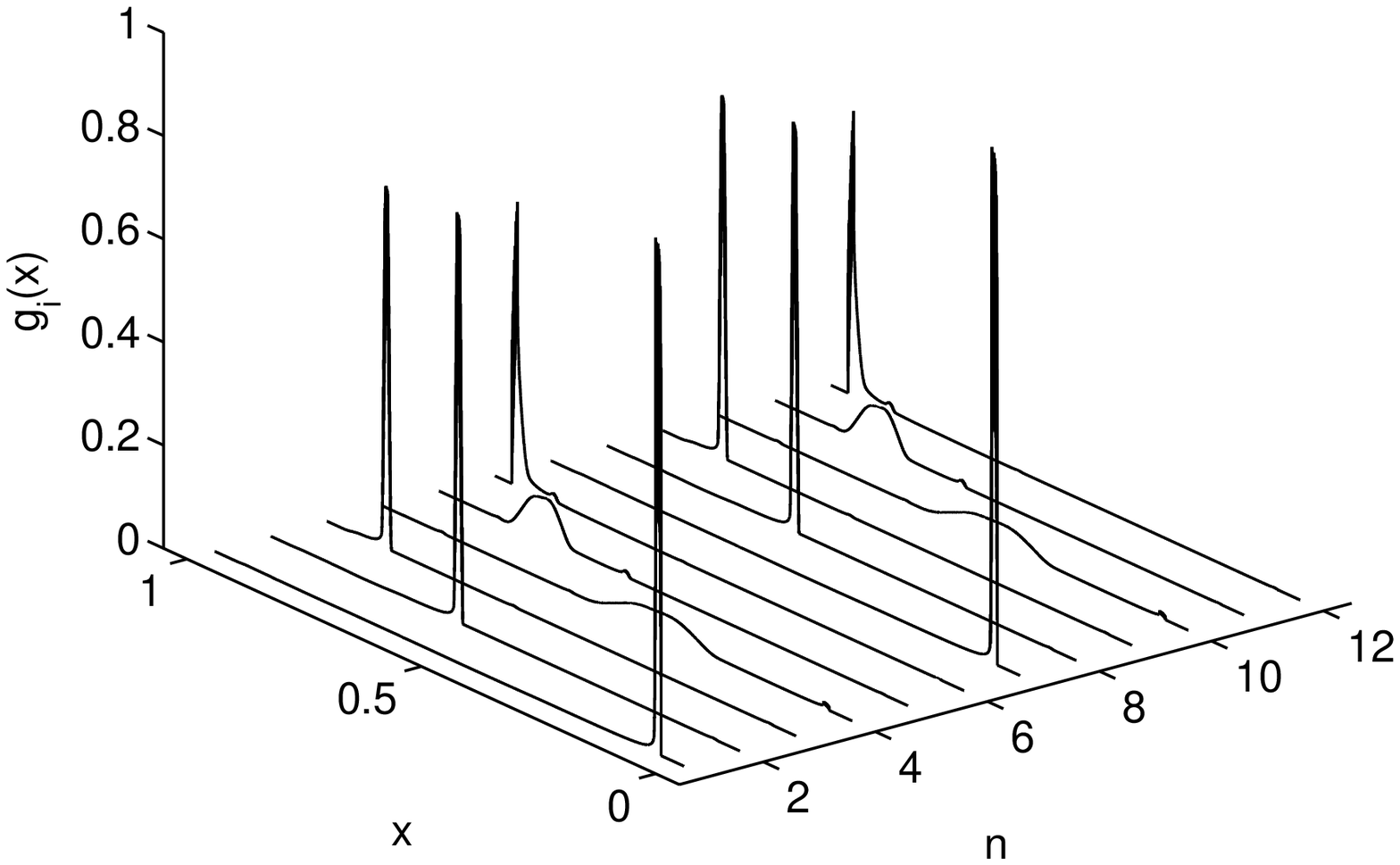}
\hspace{10mm}
\includegraphics[scale=0.3]{denstraj0.04.eps}
\includegraphics[scale=0.3]{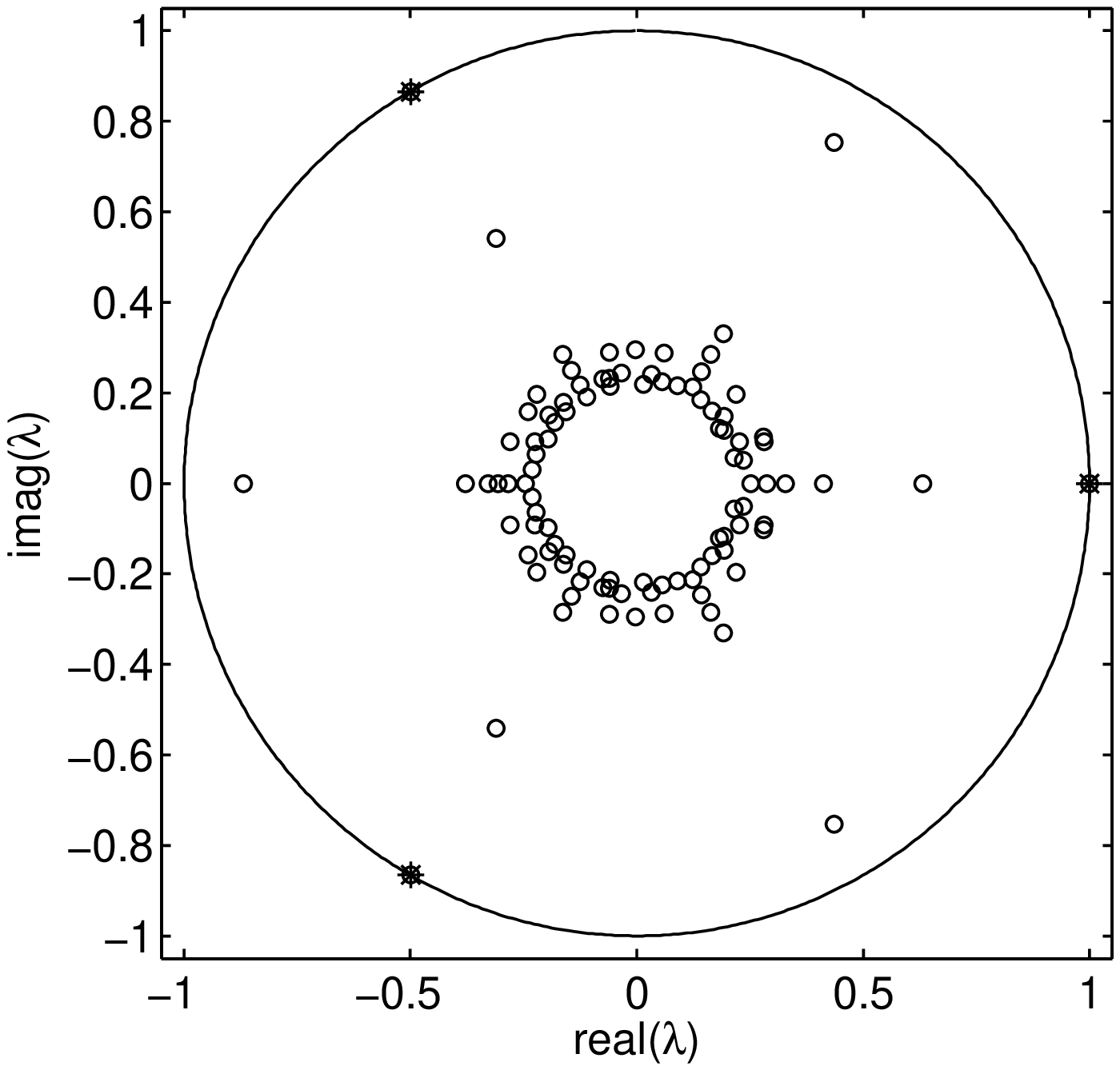}
\includegraphics[scale=0.3]{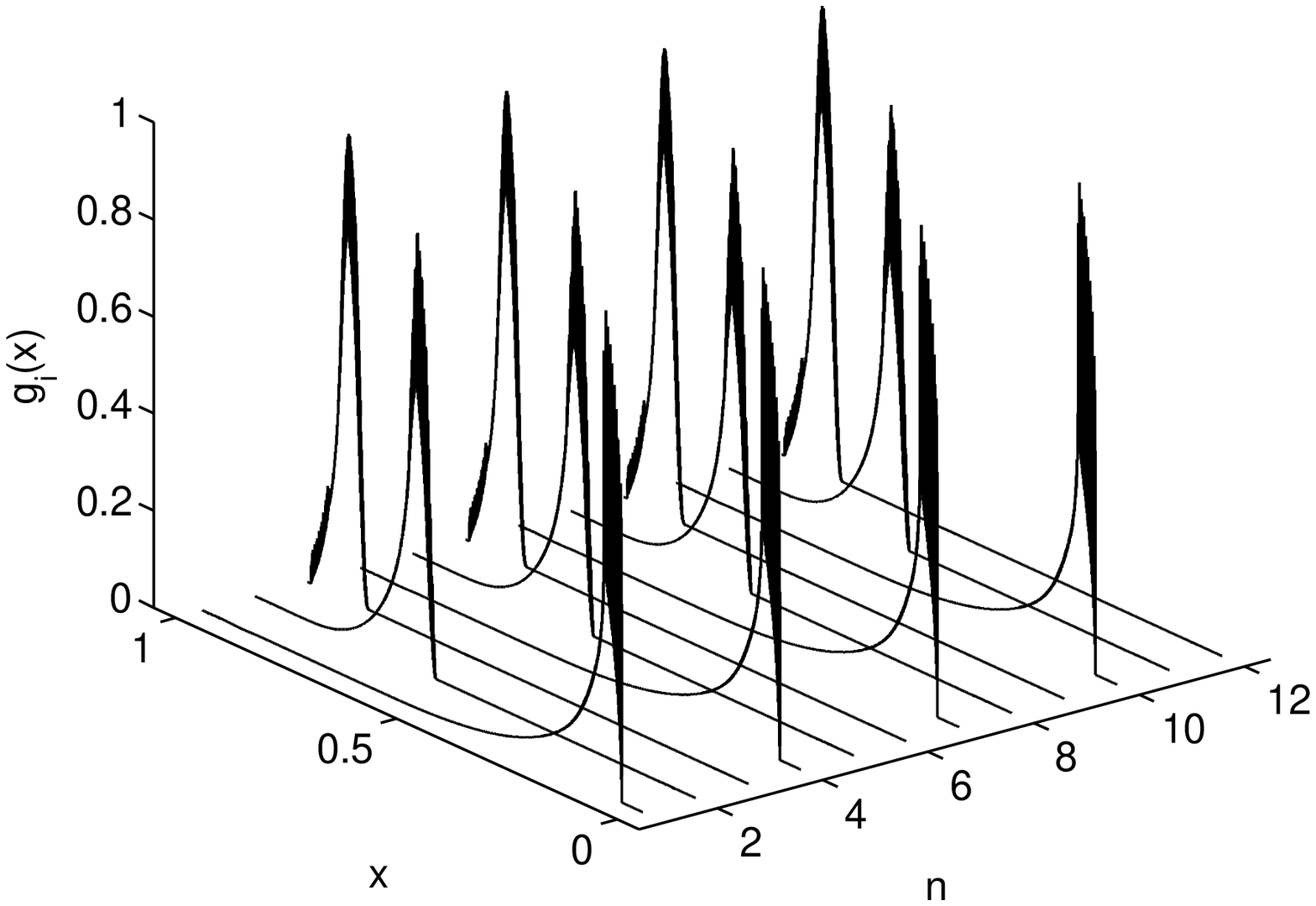}

\end{center}
\caption{Eigenvalues and eigenfunctions of the perturbed Perron-Frobenius operator: The parameters are $\alpha=1/2, \lambda=17/30, \beta=120$. The noise amplitude $\theta$ is varied $\theta=0.012$ (top) and $\theta=0.04$ (bottom). In the figures, densities are approximated as histograms with intervals with width $10^{-3}$ made from $10^6$ sample trajectories (left). Statistical periodicity of period $6$ is observed for $\theta=0.012$ (top). The supports of six almost cyclic sets are described by the leading eigenfunctions $g_i(x) ~(i=1,2,3,4,5,6)$ (top right) of the transfer operator as discussed in \cite{dellnitz99}. The corresponding top $6$ eigenvalues are given by $\gamma_1=1, \gamma_2=-0.4959+0.8611i, \gamma_3=-0.4959-0.8611i, \gamma_4=-1, \gamma_5=0.4956+0.8537i, \gamma_6=0.4956-0.8537i$, whose values are very close to the sixth roots of unity (top middle). 
Statistical periodicity of period $3$ is observed for $\theta=0.04$ (bottom) with $g_i(x) ~(i=1,2,3)$ (bottom right) and the corresponding top $3$ eigenvalues, $\gamma_1=1, \gamma_2=-0.4995+0.8655i, \gamma_3=-0.4995-0.8655i$, whose values are very close the third roots of unity (bottom middle). When $\theta=0$ and $\theta=0.4$, we have an asymptotically stable invariant densities with the unique top eigenvalue 1.}
\label{fig:density}
\end{figure}

\begin{figure}[htbp]
\begin{center}
\includegraphics[scale=0.6]{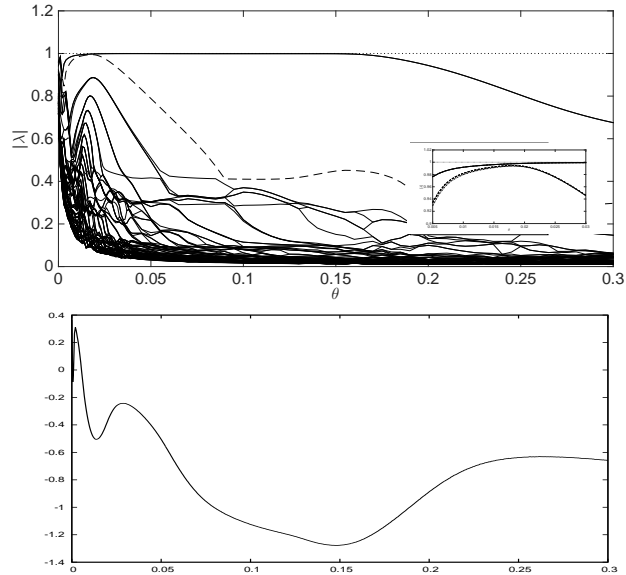}
\end{center}
\caption{Modulus of the eigenvalues of the transfer operator (top) and Lyapunov exponent (bottom) as a function of the noise intensity $\theta$.}
\label{fig:ev}
\end{figure}

Without noise, the system has an asymptotically stable invariant density and trajectories are chaotic. The density shows period $6$ and $3$ at the given coarse-grained scales $10^{-3}$ in our numerical experiments (Fig. \ref{fig:density}), while the original Lasota-Mackey random map shows only period $3$ with the same parameter. The supports of the leading eigenfunctions $g_i(x) ~(i=1,2,3,\ldots)$ of the transfer operator corresponding to eigenvalues $\gamma_i ~(i=1,2,3,\ldots)$ that are very close to the third roots of unity and whose absolute values are 1 or very close to 1 describe the almost cyclic sets and thus the almost periodic behaviour of the density in finite time. 

In Fig. \ref{fig:ev} (top) the modulus of the top eigenvalues of the transfer operator is plotted in dependence of the noise intensity $\theta$. The dashed line corresponds to the family of eigenvalues close to the sixth roots of unity corresponding to the almost six-cyclic behaviour. This dynamics is only observable only for a small range of small $\theta$, when the absolute value of these eigenvalues is very close to one.  The three-cyclic dynamics is taking over at about $\theta=0.02$ (see also inset) and then dominates until the corresponding eigenvalues go down (at about $\theta=0.17$) and the asymptotic stability of the invariant density is also observed in finite time. 

Noise-induced statistical periodicity relates to noise-induced order, which is known as a noise-induced phenomenon emerging from chaotic attracter under presence of external noise. Starting with chaotic behaviour with positive Lyapunov exponents, adding small noise induces ordered behaviour with negative Lyapunov exponents. In our model in statistically periodic region, we typically observe the localisation of the density in excitable dynamics, which consists of narrow steep expanding parts and wide gently-sloping contracting parts. When statistical periodicity occurs, the physical measure in the expanding area substantially decreases and becomes less contributing to the Lyapunov exponents. The net effects results in the transition of positivity to negativity of the top Lyapunov exponents, see Fig. \ref{fig:ev} (bottom). It is reported that there are multiple characteristic frequencies emerging in noise-induced order\cite{Matsumoto83}, which correspond to the multiple characteristic periods in terms of statistical periodicity as described above (see also Fig. \ref{fig:ev} (top)).

With smaller noise, more complex multiple transitions of periodicity of density have been observed in our model. They can be analysed by a combination of spectral analysis of the perturbed transfer operators, stability analysis with Lyapunov exponents, and finite time/space Lyapunov exponents, which are left to future works.

We have suggested a class of random dynamical systems which shows noise-induced statistical periodicity, and analyse it in terms of spectral decomposition and with the concept of almost-cyclic sets. Relationship between noise-induced periodicity and noise-induced order is also explained in this framework. A more precise transition point analysis of noise-induced statistical periodicity in this model can be studied by means of validated numerics \cite{tucker2011validated}, which will be done elsewhere. 

\vspace{2mm}

\section*{Acknowledgements}
The authors thank Prof. Yoichiro Takahashi (University of Tokyo), Prof. Michiko Yuri (Hokkaido University), Prof. Takehiko Morita (Osaka University),  Prof. Hiroki Sumi (Kyoto University), and Yukiko Iwata (Tohoku Gakuin University) for meaningful discussions. YS thanks the London Mathematical Laboratory for supports. YS is supported by the JSPS Grant-in-Aid for Scientific Research (C) No. 24540390 and 18K03441. KPG acknowledges funding from the European Union's Horizon 2020 research and innovation programme under the Marie Sklodowska-Curie grant agreement No 643073.

\bibliography{nisp1dm}

\begin{thebibliography}{14}
\expandafter\ifx\csname natexlab\endcsname\relax\def\natexlab#1{#1}\fi
\expandafter\ifx\csname bibnamefont\endcsname\relax
  \def\bibnamefont#1{#1}\fi
\expandafter\ifx\csname bibfnamefont\endcsname\relax
  \def\bibfnamefont#1{#1}\fi
\expandafter\ifx\csname citenamefont\endcsname\relax
  \def\citenamefont#1{#1}\fi
\expandafter\ifx\csname url\endcsname\relax
  \def\url#1{\texttt{#1}}\fi
\expandafter\ifx\csname urlprefix\endcsname\relax\def\urlprefix{URL }\fi
\providecommand{\bibinfo}[2]{#2}
\providecommand{\eprint}[2][]{\url{#2}}

\bibitem[{\citenamefont{Kifer}(1974)}]{Kifer74}
\bibinfo{author}{\bibfnamefont{J.~I.} \bibnamefont{Kifer}},
  \bibinfo{journal}{Math. USSR-Izu.} \textbf{\bibinfo{volume}{8}},
  \bibinfo{pages}{1083} (\bibinfo{year}{1974}).

\bibitem[{\citenamefont{Mayer-Kress and Haken}(1981)}]{Mayer-Kress81}
\bibinfo{author}{\bibfnamefont{G.}~\bibnamefont{Mayer-Kress}} \bibnamefont{and}
  \bibinfo{author}{\bibfnamefont{H.}~\bibnamefont{Haken}},
  \bibinfo{journal}{Journal of Statistical Physics}
  \textbf{\bibinfo{volume}{26}}, \bibinfo{pages}{149} (\bibinfo{year}{1981}).

\bibitem[{\citenamefont{Crutchfield et~al.}(1982)\citenamefont{Crutchfield,
  Farmer, and Huberman}}]{Crutchfield82}
\bibinfo{author}{\bibfnamefont{J.~P.} \bibnamefont{Crutchfield}},
  \bibinfo{author}{\bibfnamefont{J.~D.} \bibnamefont{Farmer}},
  \bibnamefont{and} \bibinfo{author}{\bibfnamefont{B.~A.}
  \bibnamefont{Huberman}}, \bibinfo{journal}{Physics Letters}
  \textbf{\bibinfo{volume}{92:2}}, \bibinfo{pages}{45} (\bibinfo{year}{1982}).

\bibitem[{\citenamefont{Matsumoto and Tsuda}(1983)}]{Matsumoto83}
\bibinfo{author}{\bibfnamefont{K.}~\bibnamefont{Matsumoto}} \bibnamefont{and}
  \bibinfo{author}{\bibfnamefont{I.}~\bibnamefont{Tsuda}},
  \bibinfo{journal}{Journal of Statistical Physics}
  \textbf{\bibinfo{volume}{31}}, \bibinfo{pages}{87} (\bibinfo{year}{1983}).

\bibitem[{\citenamefont{Lasota and Mackey}(1987)}]{Lasota87}
\bibinfo{author}{\bibfnamefont{A.}~\bibnamefont{Lasota}} \bibnamefont{and}
  \bibinfo{author}{\bibfnamefont{M.}~\bibnamefont{Mackey}},
  \bibinfo{journal}{Physica} \textbf{\bibinfo{volume}{D 28}},
  \bibinfo{pages}{143} (\bibinfo{year}{1987}).

\bibitem[{\citenamefont{Keener}(1980)}]{Keener80}
\bibinfo{author}{\bibfnamefont{J.~P.} \bibnamefont{Keener}},
  \bibinfo{journal}{Transactions of the American Mathematical Society}
  \textbf{\bibinfo{volume}{261:2}}, \bibinfo{pages}{589}
  (\bibinfo{year}{1980}).

\bibitem[{\citenamefont{Lasota and Mackey}(1991)}]{Lasota91}
\bibinfo{author}{\bibfnamefont{A.}~\bibnamefont{Lasota}} \bibnamefont{and}
  \bibinfo{author}{\bibfnamefont{M.}~\bibnamefont{Mackey}},
  \emph{\bibinfo{title}{Chaos, Fractals, and Noise}}
  (\bibinfo{publisher}{Springer}, \bibinfo{year}{1991}).

\bibitem[{\citenamefont{Nagumo and Sato}(1972)}]{nagumo1972response}
\bibinfo{author}{\bibfnamefont{J.}~\bibnamefont{Nagumo}} \bibnamefont{and}
  \bibinfo{author}{\bibfnamefont{S.}~\bibnamefont{Sato}},
  \bibinfo{journal}{Kybernetik} \textbf{\bibinfo{volume}{10}},
  \bibinfo{pages}{155} (\bibinfo{year}{1972}).

\bibitem[{\citenamefont{Aihara et~al.}(1990)\citenamefont{Aihara, Takabe, and
  Toyoda}}]{Aihara90}
\bibinfo{author}{\bibfnamefont{K.}~\bibnamefont{Aihara}},
  \bibinfo{author}{\bibfnamefont{T.}~\bibnamefont{Takabe}}, \bibnamefont{and}
  \bibinfo{author}{\bibfnamefont{M.}~\bibnamefont{Toyoda}},
  \bibinfo{journal}{Physics Letters} \textbf{\bibinfo{volume}{A 144}},
  \bibinfo{pages}{333} (\bibinfo{year}{1990}).

\bibitem[{\citenamefont{Dellnitz and Junge}(1999)}]{dellnitz99}
\bibinfo{author}{\bibfnamefont{M.}~\bibnamefont{Dellnitz}} \bibnamefont{and}
  \bibinfo{author}{\bibfnamefont{O.}~\bibnamefont{Junge}},
  \bibinfo{journal}{SIAM Journal on Numerical Analysis}
  \textbf{\bibinfo{volume}{36}}, \bibinfo{pages}{491} (\bibinfo{year}{1999}).

\bibitem[{\citenamefont{Komornik and Lasota}(1987)}]{Komornik87}
\bibinfo{author}{\bibfnamefont{J.}~\bibnamefont{Komornik}} \bibnamefont{and}
  \bibinfo{author}{\bibfnamefont{A.}~\bibnamefont{Lasota}},
  \bibinfo{journal}{Bull. Polon. Acad. Sci. Math.}
  \textbf{\bibinfo{volume}{35}}, \bibinfo{pages}{321} (\bibinfo{year}{1987}).

\bibitem[{\citenamefont{Chekroun et~al.}(2011)\citenamefont{Chekroun, Simonnet,
  and Ghil}}]{chekroun2011stochastic}
\bibinfo{author}{\bibfnamefont{M.~D.} \bibnamefont{Chekroun}},
  \bibinfo{author}{\bibfnamefont{E.}~\bibnamefont{Simonnet}}, \bibnamefont{and}
  \bibinfo{author}{\bibfnamefont{M.}~\bibnamefont{Ghil}},
  \bibinfo{journal}{Physica D: Nonlinear Phenomena}
  \textbf{\bibinfo{volume}{240}}, \bibinfo{pages}{1685} (\bibinfo{year}{2011}).

\bibitem[{\citenamefont{Faranda et~al.}(2017)\citenamefont{Faranda, Sato,
  Saint-Michel, Wiertel, Padilla, Dubrulle, and
  Daviaud}}]{faranda2017stochastic}
\bibinfo{author}{\bibfnamefont{D.}~\bibnamefont{Faranda}},
  \bibinfo{author}{\bibfnamefont{Y.}~\bibnamefont{Sato}},
  \bibinfo{author}{\bibfnamefont{B.}~\bibnamefont{Saint-Michel}},
  \bibinfo{author}{\bibfnamefont{C.}~\bibnamefont{Wiertel}},
  \bibinfo{author}{\bibfnamefont{V.}~\bibnamefont{Padilla}},
  \bibinfo{author}{\bibfnamefont{B.}~\bibnamefont{Dubrulle}}, \bibnamefont{and}
  \bibinfo{author}{\bibfnamefont{F.}~\bibnamefont{Daviaud}},
  \bibinfo{journal}{Physical review letters} \textbf{\bibinfo{volume}{119}},
  \bibinfo{pages}{014502} (\bibinfo{year}{2017}).

\bibitem[{\citenamefont{Tucker}(2011)}]{tucker2011validated}
\bibinfo{author}{\bibfnamefont{W.}~\bibnamefont{Tucker}},
  \emph{\bibinfo{title}{Validated numerics: a short introduction to rigorous
  computations}} (\bibinfo{publisher}{Princeton University Press},
  \bibinfo{year}{2011}).

\end{thebibliography}

\end{document}